\documentclass[aps,floats,superscriptaddress,showpacs]{revtex4}
\usepackage{amsmath}

\newcommand{\beq}{\begin{equation}}
\newcommand{\eeq}{\end{equation}}
\newcommand{\beqa}{\begin{eqnarray}}
\newcommand{\eeqa}{\end{eqnarray}}
\usepackage{epsfig}
\usepackage{graphicx}% Include figure files
\usepackage{dcolumn}% Align table columns on decimal point
\usepackage{bm}% bold math

\def\gapp{\lower.35em\hbox{$\stackrel{\textstyle>}{\sim}$}}
\def\lapp{\lower.35em\hbox{$\stackrel{\textstyle<}{\sim}$}}

\begin{document}
\bibliographystyle{apsrev}
%

%\draft
\title{Renormalization group  aspects of graphene}

\author{ Mar\'{\i}a A. H. Vozmediano}

\affiliation{Instituto de Ciencia de Materiales de Madrid,\\
CSIC, Cantoblanco, E-28049 Madrid, Spain.}

\date{\today}
%%%%%%%%%%%%%%%%%%%%%%%%%%%%%%%%%%%%%%%%%%%%%%%%%%%%%%%%%%%%%%%%%%%%%%%%%%%%%
\begin{abstract}
Graphene is a two dimensional crystal of carbon atoms with fascinating
electronic and morphological properties. The low energy excitations of the
neutral, clean system are  described by a massless Dirac Hamiltonian 
in (2+1) dimensions which also captures the main electronic and transport properties. 
A renormalization group analysis  sheds light on the success of the free model:
due to the special form of the Fermi surface which reduces to two single points in 
momentum space, short range interactions are irrelevant and  only gauge
interactions like long range Coulomb or effective disorder can play 
a role in the low energy physics. We  review these features and discuss briefly
other aspects related to disorder and to the bilayer material along the same lines.

\end{abstract}
%%%%%%%%%%%%%%%%%%%%%%%%%%%%%%%%%%%%%%%%%%%%%%%%%%%%%%%%%%%%%%%%%%%%%%%%%%%%%
%
%\pacs{75.10.Jm, 75.10.Lp, 75.30.Ds}
%
%
%%%%%%%%%%%%%%%%%%%%%%%%%%%%%%%%%%%%%%%%%%%%%%%%%%%%%%%%%%%%%%%%%%%%%%%%%%%%%
%%%%%%%%%%%%%%%%%%%%%%%%%%%%%%%%%%%%%%%%%%%%%%%%%%%%%%%%%%%%%%%%%%%%%%%%%%%%%
%%%%%%%%%%%%%%%%%%%%%%%%%%%%%%%%%%%%%%%%%%%%%%%%%%%%%%%%%%%%%%%%%%%%%%%%%%%%%
%%%%%%%%%%%%%%%%%%%%%%%%%%%%%%%%%%%%%%%%%%%%%%%%%%%%%%%%%%%%%%%%%%%%%%%%%%%%%
%

\maketitle
 \section{Introduction.}
 \label{intro}
For the last few decades, the Landau Fermi liquid (LFL) has been the standard model in condensed matter to describe most  metals \cite{L57,Aetal75}. A renormalization group (RG) analysis of the effective continuum model of non-relativistic electrons allows to understand this universal behavior as a consequence of the marginal character of electron-electron interactions in the presence of a Fermi surface \footnote{See the article by R. Shankar in this volume.}. Strongly correlated materials such as heavy fermions or the hight temperature superconductors have given rise to very important new ideas in the last 20 years due to their ``non--Fermi liquid" behavior that arise from their strongly interacting nature \cite{QH09}. The appearance of graphene in 2004 \cite{Netal04} has originated a ``second revolution" in the field due to its very many exotic electronic and morphological properties. Unlike the cuprates where there is not up to today an agreement on the model to describe the basic material, graphene resembles the situation in quantum field theory (QFT) where the starting point is clear: the low energy excitations can be modeled  by the massless Dirac Hamiltonian in two spatial dimensions. Under the RG point of view, graphene has defied the LFL paradigm by confronting us with a material described by a universal non-interacting Hamiltonian  that shows agreements and contradictions with the usual LFL behavior. From many other points of view it is a fascinating system which lives in between different branches of physics as condensed matter, QFT, statistical physics or the theory of elasticity.

Graphene is a two-dimensional crystal of carbon atoms arranged in a honeycomb lattice: a single layer of graphite. Its synthesis \cite{Netal05,Zetal05}, amazing properties ({\it it is flexible like plastic but stronger than diamond, and it conducts electricity like a metal but is transparent like glass} \footnote{Quoted from Science NOW, October 2010}) and potential applications \cite{G09,SMD10}, have  granted the 2010 Nobel prize in Physics  to their fathers A. Geim and K. Novoselov. 

One of the most interesting aspects of the graphene physics from a theoretical point of view is the deep and fruitful relation that it has with quantum electrodynamics (QED) and other quantum field theory ideas \cite{S84,H88,GGV94,KN07}. The connection arises from the mentioned  fact that the low energy excitations can be modeled  by the massless Dirac equation in two spatial dimensions a fact that makes useless most of the phenomenological expressions for transport properties in Fermi liquids. From the morphological point of view, the mere existence of two dimensional crystals has been argued not to be possible \cite{G09}. Moreover despite of being one of the most rigid materials with a Young modulus of the order of terapascals, the samples show ripples of various sizes whose origin is unknown.   The intrinsic relation between the morphology (honeycomb lattice) and the electronic properties constitutes a great playground for physical ideas as well as one of the sources of potential applications as a single molecule detector.

Under the QFT point graphene has given rise to  very interesting developments: the so-called axial anomaly \cite{S84,H88} has acquired special relevance in relation with the recently discussed topological insulators \cite{BHZ06,HK10} which provide a condensed matter realization of the axion electrodynamics \cite{W87,LWetal10}. Charge fractionalization has also been explored in the honeycomb lattice with special defects \cite{HCM07,JP07} and quantum field theory in curved space \cite{GGV92,JCV07} and cosmological models  \cite{CV07a,CV07b} have been used to explore the electronic properties of the curved material. A very interesting development is associated to the generation of various types of vector fields coming from the elastic properties or from disorder that couple to the electrons in the form of gauge fields \cite{VKG10}.

Among the unexpected properties related with interactions on the condensed matter point of view are the anomalous behavior of the quasiparticles decaying linearly with frequency \cite{LLA09}, and the renormalization of the Fermi velocity at low energies \cite{Letal08}. These properties were predicted theoretically in the early works \cite{GGV94,GGV96} from the RG analysis that will be described in this work. We will review the  RG aspects of graphene in the presence of Coulomb interactions with special emphasis in the similarities and discrepancies with QED(3).  We will show that the system has a non--trivial infrared fixed point where Lorentz covariance is restored appearing as an emergent property \cite{V03}. The RG analysis allows to classify graphene as a new type of electron liquid described by a Lorentz covariant infrared fixed point whose effective coupling constant is the fine structure constant. We will focus on the general, conceptual features referring to the specific
bibliography for the technical details. The differences and similarities with the usual QFT models will also be emphasized. %The work is organized as follows: Section \ref{sec_model} describes the continuum model for the electrons in the carbon honeycomb lattice. Section \ref{sec_qft} etc- 

\section{Continuum model for the low energy  excitations of graphene}
\label{sec_model}
\begin{figure}
\begin{center}
\includegraphics[height=4cm]{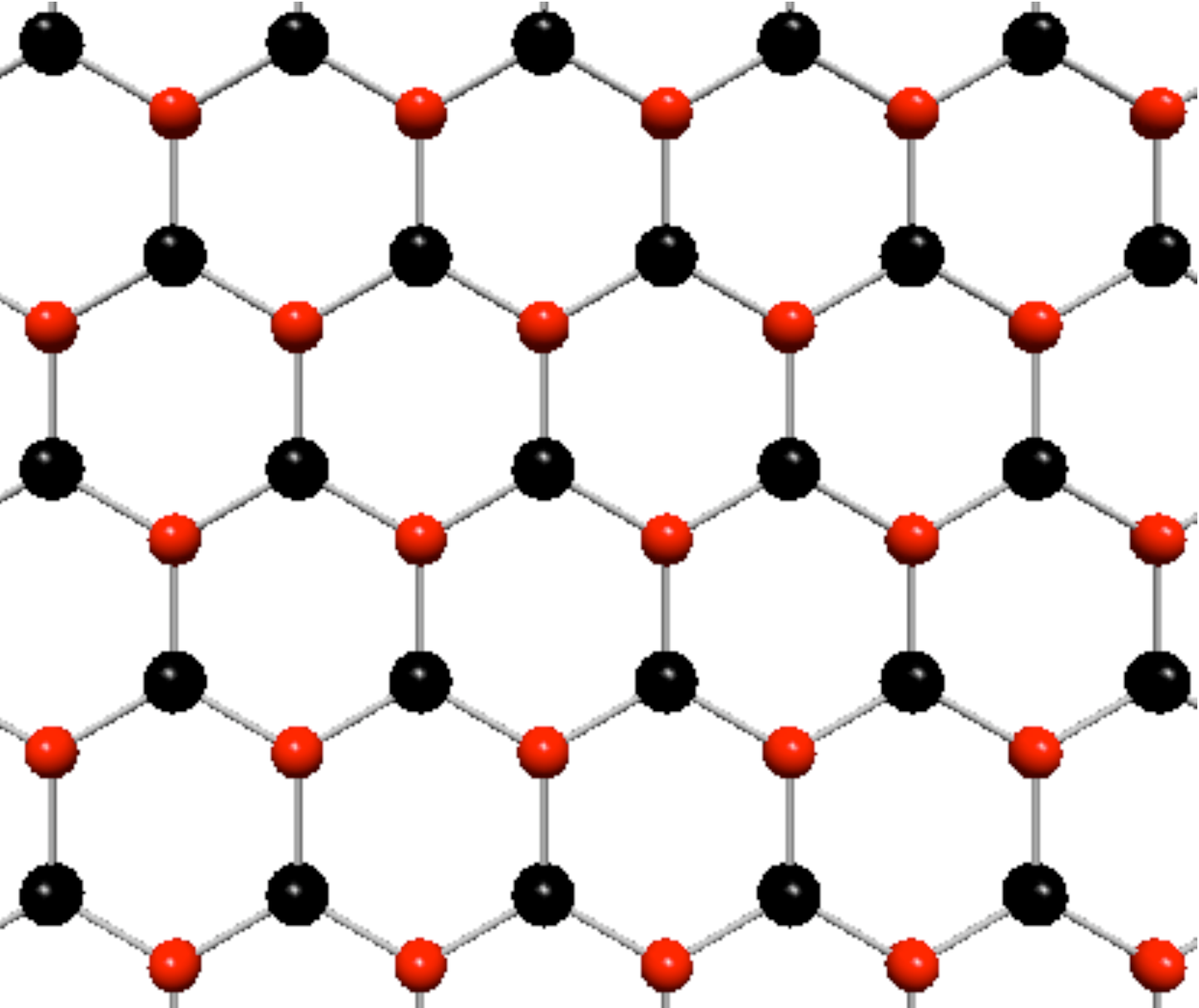}
\caption{(Color online) Honeycomb lattice of monolayer graphene. }
\label{lattice}
\end{center}
\end{figure}
The graphene structure is made of carbon atoms located in a honeycomb lattice. Three of the four available electrons of the carbon atom: 2$s$, 2$p_x$ and 2$p_y$ orbitals hybridize in a so-called $sp^2$, a strong covalent bond that "draw" the honeycomb lattice and have  typical energies around 3 eV. The fourth  2$p_z$ orbital perpendicular to the plane formed by the sigma bonds remains delocalized and is responsible for the metallic properties of the material. The $\sigma$ bonds give rigidity to the structure, while the $\pi$ bonds give rise to the valence
and conduction bands \cite{SDD98}. The electronic properties around the Fermi
energy of a graphene sheet can be described by a tight binding
model with only one orbital per atom, the so-called $\pi$-electron
approximation and were obtained in the early works \cite{W47,SW58}.
The nearest-neighbor tight binding approach reduces the problem to the
diagonalization of the one-electron Hamiltonian

\begin{align}
{\cal H} =  -t \sum_{<i,j>} a^+ _i a_j
\label{oneehamil}
\end{align}
where the sum is over pairs of nearest neighbors atoms $i,j$ on the lattice
and $a_i$, $a^+_j$ are the usual creation and annihilation operators.
What is special about the honeycomb lattice shown in Fig. \ref{lattice}
is the fact that, even though all carbon atoms are equal, their positions in the lattice
makes them topologically non-equivalent. It is easy to see that to generate all the points in the lattice we must take a pair of (red and black) points. The lattice can be seen as two interpenetrated triangular sublattices of red and black points or, more technically, it is a lattice with a basis. This special topology of the honeycomb lattice is at the heart of all the electronic peculiarities of graphene. The Bloch trial wave function has to be build as a superposition  of the atomic orbitals from
the two atoms forming the primitive cell black (A) and red (B):  
\beq
\Psi_{\bf k}({\bf r})=C_A\phi_A+C_B\phi_B.
\eeq

\begin{figure}
\begin{center}
\includegraphics[height=6cm]{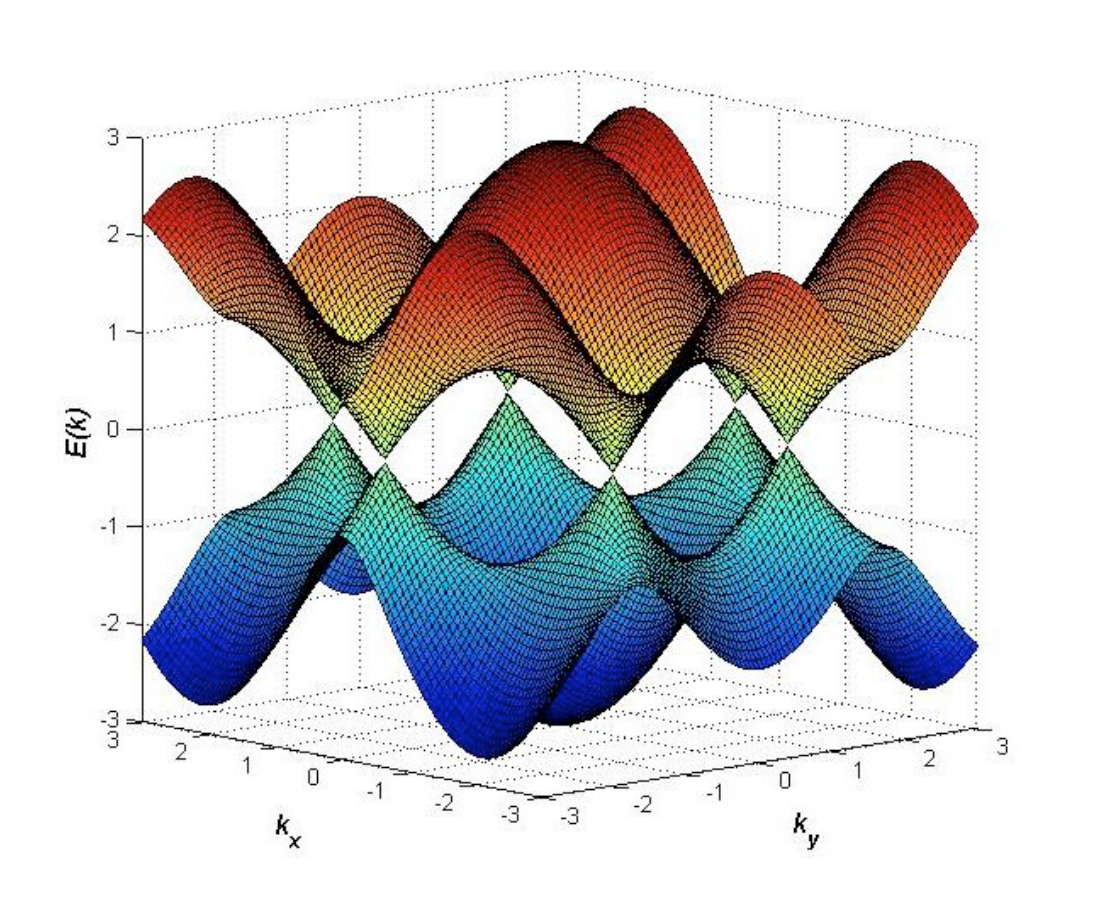}
\caption{(Color online) The dispersion relation of graphene. The six Fermi points of the
neutral material are located at the vertices of the hexagonal Brillouin zone. }
\label{disp}
\end{center}
\end{figure}

The eigenfunctions and eigenvalues of the Hamiltonian are obtained from
the equation
\begin{eqnarray}
\left(
\begin{array}{cc} \epsilon & -t\sum_j e^{ia{\bf k·u_j}}
\\ -t\sum_j e^{ia{\bf k·v_j}} &\epsilon \end{array} \right)
\left( \begin{array}{cc} C_A \\ C_B \end{array} \right)
= E( {\bf k }) \left( \begin{array}{cc} C_A \\ C_B \end{array} \right)\;,
\label{eigen}
\end{eqnarray}
where ${\bf u_j}$ is a triad of vectors connecting an
A atom with its B nearest neighbors
and ${\bf v_j}$ the triad of their respective opposites,
$a$ is the distance between carbon
atoms and $\epsilon$ is the 2$p_z$ energy level,
taken as the origin of the energy. The tight binding parameter $t$  estimated to be of the 
order of 3eV in graphene sets the bandwidth (6eV) and is 
a measure of the kinetic energy of the electrons. 
The eigenfunctions are determined by the the coefficients
$C_A$ and $C_B$ solutions of equation (\ref{eigen}). The
eigenvalues of the equation  give the energy levels whose dispersion relation is
\begin{align}
E({\bf k})=\pm t\sqrt{1+4\cos^2\frac{\sqrt{3}}{2}ak_x+
4\cos\frac{\sqrt{3}}{2}ak_x\cos\frac{3}{2}ak_y} \;,
\label{disprel}
\end{align}
in which the two signs define two energy bands: the lower half called
the bonding $\pi$ band and the upper half called the
antibonding $\pi^*$ bands, which
are degenerate at the $ {\bf K}$  points of the Brillouin zone.
The dispersion relation is shown in Figure \ref{disp}.
Within the $\pi$ electron approximation each site of the
graphite honeycomb lattice yields one electron
to the Fermi sea and the band is at half-filling.
Since each level of the band may accommodate two
states due to the spin degeneracy, and the Fermi
level turns out to be at the midpoint of the band,
instead a whole Fermi line,  the 2D honeycomb lattice has six isolated Fermi
points which are the six vertices of the hexagonal
Brillouin zone. Only two of them
are inequivalent and can be chosen as $\vec K_1 =(4\pi/3\sqrt{3},0)
$ and $\vec K_2= -\vec K_1$ \cite{GGV93}.

A continuum model can be defined for the low energy excitations 
around any of the Fermi points, say $K_1$,
by expanding the dispersion relation around it:
${\bf k}=K_1+{\bf \delta k}$ what gives from (\ref{eigen}) the Hamiltonian

\begin{eqnarray}
{\cal H}\sim -\frac{3}{2}ta \left(
\begin{array}{cc} 0 & \delta k_x+i\delta k_y
\\ \delta k_x-i\delta k_y  & 0 \end{array} \right)
.
\label{contin}
\end{eqnarray}

The limit $\lim_{a\to 0}{\cal H}/a$ defines the continuum Hamiltonian
\beq
H=v_F {\vec \sigma} .{\vec k},
\label{dirach}
\eeq
where $\sigma$ are the Pauli matrices and the parameter $v_F$ is the Fermi velocity of the
electrons estimated to be $v_F\sim 3ta\sim c/300$. Hence the low energy excitations of the system 
are massless, charged spinors in two spatial dimensions moving at a speed $v_F$. We must notice that the physical spin of the electrons have been neglected in the analysis, the spinorial nature of the wave function has its origin in the sublattice degrees of freedom and is called pseudospin in the graphene literature. In the absence of interaction mixing the two Fermi points it can be shown that they are topologically stable \cite{MGV07} and can not be lifted by smooth continuum  perturbations of the Hamiltonian, another way of seeing the RG results that will be analyzed here. The same expansion around the other Fermi point gives rise to a time reversed Hamiltonian: ${\cal H}_2=v_F(-\sigma_x k_x+\sigma_y k_y)$. The degeneracy associated to the Fermi points (valleys in the semiconductor's language) is taken as a flavor. Together with the real spin the total degeneracy of the system is 4.

The calculation of the density of states for the linear dispersion relation $E (k)=v_F\vert k \vert$ gives
\begin{equation}
\rho (E)=\frac{g_{S}}{ v_{F}^{2}} \frac{\left|E\right|}{2\pi},
\label{dos}
\end{equation}
where $g_{S}$=4 is the degeneracy.  Eq. (\ref{dos})  
shows one of the peculiarities of graphene as opposed to a usual two dimensional electron gas (2DEG): the density of states grows linearly with the energy and vanishes at the Fermi energy (in the standard 2DEG the DOS is a constant in two dimensions). This fact has very important phenomenological consequences that have been tested experimentally. For our purposes it implies that the Coulomb interactions between the electrons is not screened in graphene. 

We finish this section by mentioning that the massless spinorial nature of the graphene quasiparticles, the linearity of the DOS, the degeneracy of 4 and the estimated value of the Fermi velocity have been experimentally tested \cite{G09} what makes the present model universally accepted in the community. We also mention that many  electronic and transport properties can be described by the non-interacting model and that disorder which usually leads to localization in two dimensions, does not seem to play an important role in the graphene system.

\section{RG analysis: tree level}
\label{sec_rg}
As discussed in the article by R. Shankar in this volume, the main issue of the renormalization group ideas as applied to condensed matter systems is
that for special systems (critical, renormalizable) the low energy
physics is governed by an effective Hamiltonian
made of a few marginal interactions that can be obtained
from the microscopic high-energy Hamiltonian in a well
prescribed manner \cite{P93,S94}.
Following the nomenclature of  critical systems, interactions are classified as relevant, irrelevant or
marginal according to their scale dimensions. These dimensions
determine whether they grow, decrease, or acquire
at most logarithmic corrections at low energies.
The effective coupling constants of a model at intermediate
energies can be obtained by ``integrating out" high energy
modes even if there is no stable fixed point at the
end of the RG flow. The Luttinger and Fermi liquids are
identified as infrared fixed points of the RG applied to
an interacting metallic system in one or more dimensions
respectively.

In this context, the universal Landau Fermi liquid 
behavior of most electronic systems can be established in the form of 
almost {\it a theorem}: a system of electrons will behave as a Fermi liquid if: (1) The spatial dimension is $D\geq2$.
(2) The system has a smooth extended Fermi surface.  
(3) The interactions are repulsive, non-singular and short ranged.

The key ingredient in the result is the existence of a finite Fermi surface which 
\begin{itemize}
\item Ensures a finite density of states at the Fermi energy and hence the screening of the electron-electron interaction making it effectively short ranged.
\item Makes the tree level scaling analysis effectively one dimensional as only  the momentum component perpendicular to the Fermi surface scales like the energy. 
\item Provides the kinematics necessary for the Landau channels to work.

\end{itemize}
Most of the non-Fermi liquid behaviors searched for in two dimensional systems in the last years have been associated to either specific shapes of the Fermi surface as in the so--called Van Hove scenario \cite{M98} or to singular interactions \cite{VNS02}. The case of graphene is specially interesting for the reduction of the Fermi line to two single points. This fact prompted the early RG analysis\cite{GGV94} which at the time aimed to find non-Fermi liquid behavior coming from the fact that Fermi points were characteristics of one dimensional systems with different universal behavior (Luttinger liquid). 

Next we will classify the local interactions in the graphene system at tree level system following \cite{P93,S94}. From the  Hamiltonian (\ref{dirach}) we can construct the action
\beq
S_0=\int d^2 k \; d \omega \;\bar\Psi \left[(\omega I-v_F \sigma . {\bf k}\right]\Psi,
\label{freeact}
\eeq
what allows to fix the scale dimension of the fields to $[\Psi]=-2$ in units of energy.
Notice that for usual non relativistic fermions the dispersion relation is also linearized around the Fermi surface. The crucial difference between the two cases is that, because the Fermi surface is a point, there is an isotropic scaling of the momentum so that under a rescaling of the energy $\omega\sim s\omega$, $d^2 k\sim s^2 k$. In the presence of an extended Fermi line, we would have 
$d^2 k=d k_\parallel \;d k_\perp \sim d k_\parallel \; s d k_\perp$ and the dimension of the fields would be $[\Psi]=-3/2$ in any number of dimensions. 

Having established the different scale dimension of the fields it is obvious that four (or more) Fermi interactions - i. e. all local interactions in the condensed matter approach - are irrelevant. (Notice that they were already generically irrelevant also in the non-relativistic case and only the Landau channels corresponding to special kinematics around the Fermi line were made marginal). Since we can not find any marginal local interactions in the graphene case this finishes the analysis. No  magnetic or superconducting instabilities driven from local interactions  are to be expected in the clean graphene system assumed that the couplings are small and perturbation theory applies, a fact that also seems to be confirmed by most experiments. 

Although the only possible local interactions in non-relativistic electrons are made of polynomials in the fields we know that relativistic fermions interact through gauge fields and these interactions will be marginal. We will describe how they arise in the graphene system in the next section.
 
\section{Coulomb interaction. Graphene versus QED}

The unscreened Coulomb interaction
is usually written  as
\begin{equation}
{\cal H}_{int} = \frac{e^2}{2 \pi} \int d^2 {\bf r_1} \int d^2
{\bf r_2} \frac{\bar{\Psi} ( \bf{r_1} ) \Psi ( \bf{r_1} )
\bar{\Psi} (\bf{r_2} ) \Psi ( \bf{r_2} )} {| \bf{r}_1 - \bf{r}_2
|}.
 \label{hint}
\end{equation}
The non-local interaction (\ref{hint}) can be understood as an
effective interaction coming from  a standard gauge field coupled
to the fermions through the minimal coupling prescription giving rise to the interacting action:
\begin{equation}
S_{int}=g\int d^2x\; dt \;J^\mu (x,t)A_\mu (x,t)\;,
\label{sint}
\end{equation}
where $g=e^2/4\pi v_F$ is the dimensionless coupling constant, 
the electronic current is defined as
\beq
J^\mu =(\overline\Psi I \Psi,v_F \overline\Psi\sigma^i\Psi)\;.
\label{current}
\eeq
Notice that although the electrons are confined to live in the two-dimensional surface, the Coulomb interaction among them lives in the three dimensional space. We then face the problem of coupling the two-dimensional current to a three dimensional gauge potential. 
In ref.
\cite{GGV94}  this difficulty was solved by 
integrating the usual photon propagator over the $z$ component 
what produces an effective
two-dimensional propagator for the gauge field with a  $1/\vert {\bf
k}\vert$ dependence.  Since by construction the vector potential scales like the derivative,
the interaction (\ref{sint}) is marginal. The full action 
\begin{equation}
S=g\int d^2x\; dt \bar{\Psi}\gamma^\mu\left[\partial_\mu+ig\;A_\mu \right]\Psi\;, \qquad 
\mu=0,1,2
\label{sfull}
\end{equation}
is scale invariant. $\gamma^\mu$ are a set Dirac matrices constructed with the Pauli matrices.  The action (\ref{sfull}) looks like that of (non-relativistic) quantum
electrodynamics in two spatial dimensions but there are important differences that affect the discussion of the one loop renormalization of the coupling. It is
the anomalous photon propagator  what makes the interacting graphene system different from
QED(2+1) and the reason why the results found there can not be translated
directly to the physics of graphene. The coupling constant of QED (2+1)  has  dimension of mass (${\sqrt M}$) and the theory is called ``superrenormalizable". It means that it has less divergences than its four dimensional counterpart. In fact, there are no ultraviolet infinities in QED(2+1). By construction, the model given by (\ref{sfull}) is fully scale invariant, the coupling constant is dimensionless and, in this sense, the system resembles more what happens in QED(3+1) which is strictly renormalizable - or at a critical point- . This is also what induces a renormalization of the electron self-energy in the graphene system not present in planar QED.

\section{RG at the one loop level}
\label{sec_ren}
Once we have identified the marginal character of the Coulomb -- and in general any generalized gauge--type  interaction of the electronic current (\ref{current}) with a vector field of weight 1 --  at tree level we have to establish the behavior of the renormalized
coupling constant. 

From the structure of the
perturbative series we see that the effective -dimensionless-
coupling constant is $$ g=\frac{e^2}{4\pi v_F}$$ where e is the
electron charge and $v_F$ is the Fermi velocity. This is the equivalent of the graphene fine structure constant where $v_F$ replaces the speed of light $c$. We will proceed as in the Fermi liquid analysis and perform a perturbative renormalization at the one loop level. This amounts to  assume for the time
being that the effective coupling constant is small, something which is not guaranteed
in the suspended samples. We will comment on that later.
(The bare value of this constant at energies around 1eV is of the order of 2-3, corresponding to 300$\alpha_{QED}$).
The RG analysis of  the model (\ref{sfull}) was performed in
full detail in refs. \cite{GGV94,GGV96,GGV99,GGV01}. We will here summarize
the results obtained there.

\begin{figure}
\begin{center}
\includegraphics[height=3cm]{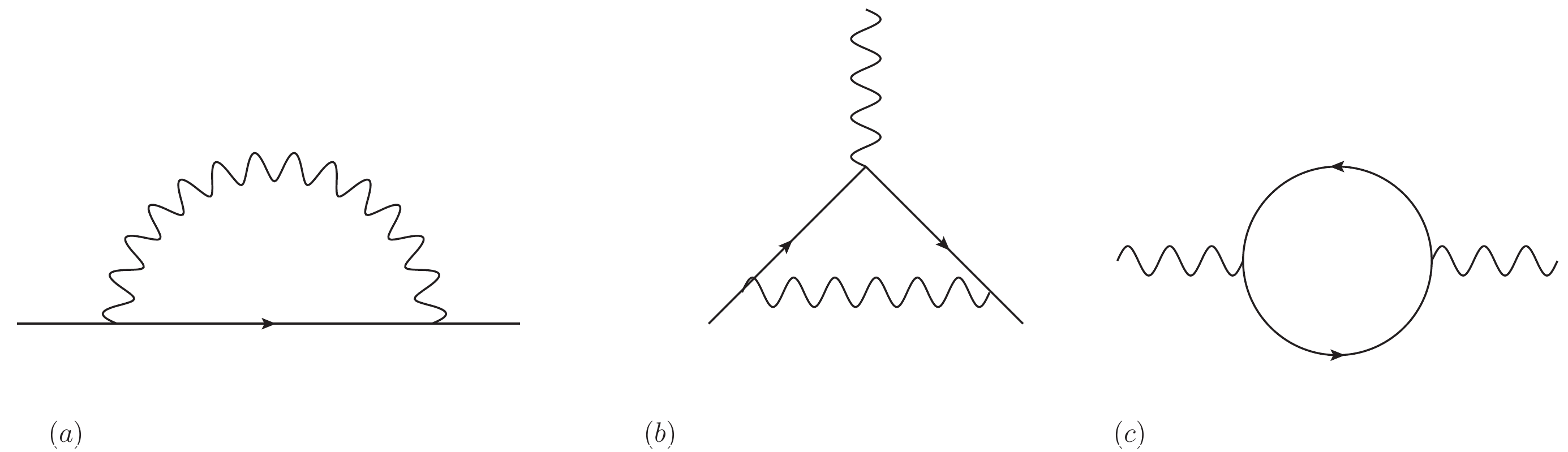}
\caption{Diagrams renormalyzing the model at one loop}
\label{oneloop}
\end{center}
\end{figure}

Due to the very special structure of the model discussed in the previous section, the coupling constant renormalization of the model at one loop level is more subtle than both cases, the non-relativistic condensed mater fermions, and QED. The building blocks of the Feynman graphs are the electron and photon propagators: 

\begin{equation}
G_0(\omega, {\bf k})=i\frac{\gamma^0\omega + v {\vec \gamma} \cdot {\bf
k}}{-\omega^2+v^2{\bf k}^2}\;,
\label{elprop}
\end{equation}
\begin{equation}
\Pi_0( {\bf k})=\dfrac{1}{2}\frac{1}{\sqrt{-\omega^2+ k^2}}\;.
\label{barephoton}
\end{equation}
and the tree level vertex
$\Gamma^\mu= ie (\gamma^0, v\;{\vec\gamma})$.

Due to the non--covariant form of the electron propagator (\ref{elprop}) the system has four free para meters, the electric charge $e$, the Fermi velocity $v$, and the electron and photon wave function renormalization. A Ward identity associated to charge conservation relates the electron propagator and the vertex function so the renormalization of the model can be done as in QED by renormalizing the electron and photon propagators only. 

The one loop diagrams are shown in Fig. \ref{oneloop}. It can be shown to all orders in perturbation theory that the photon propagator is finite and hence the electric charge is not renormalized. All the renormalization of the system comes from the electron propagator. The renormalization of the Fermi velocity and the electron  wave function are obtained  from the electron self-energy $\Sigma$ computed from  Fig. \ref{oneloop} (a) and from the relations:
\begin{equation}
\Sigma(\omega,{\bf k})=Z_\psi(\omega,{\bf k})
\left[\omega\gamma^0-Z_v(\omega,{\bf k})v\bm{\gamma} \cdot {\bf k}\right]\;.
\label{zelectron}
\end{equation}
\beq
v_0=Z_v v_R.
\eeq
The wave function renormalization
\beq
Z_\Psi\sim\frac{\partial \Sigma(\omega, {\bf
k})}{\partial\omega}\vert_{\omega=0}\;,
\eeq
defines the anomalous
dimension of the field 
\beq
\gamma=\partial \log Z_\Psi/\partial l,
\eeq
($l$ is the RG parameter) and, hence the asymptotic behavior of the fermion propagator:
\beq
G(\omega, {\bf k})\sim_{\omega\to 0}\frac{1}{\omega^{1-\eta}}\;.
\eeq
The computation gives
\beq
Z_v=1-\frac{g}{4}\log\Lambda,
\eeq
from where we can relate the Fermi velocity at two different energies:
\beq
v(E)=v(E_0)\left[1-\dfrac{g}{4}\log(\frac{E}{E_0} )\right].
\eeq
The last equation means that the Fermi velocity grows as the energy decreases, the main result of the analysis.
The anomalous dimension is found to be
\beq
\eta=\frac{e^2}{12\pi^2}.
\eeq

Summarizing, the results of \cite{GGV94} are the following:

\begin{itemize}
\item The Fermi velocity grows as the energy is decreased. This result implies
a breakdown of the relation between the energy and momentum scaling, a signature of a
quantum critical point.
\item The electron-photon vertex and the photon propagator are not
renormalized at the one loop level. This means that the electric
charge is not renormalized, a result that could be predicted by
gauge invariance, and it also implies that the effective coupling
constant $g=e^2/4\pi v_F$  decreases at low energies defining
an infrared free fixed point of the RG. 
\item The beta function of the Fermi velocity has a non--trivial zero at the value $v=c$. Hence the infrared fixed point defines a weak coupling model ruled by the fine structure constant of QED which is  Lorentz covariant. 
\item    There
is a critical exponent which determines the universality class of
the model. The anomalous exponent means that the fixed point is not a Fermi liquid and resembles more a Luttinger liquid.
\end{itemize}

\section{Marginal Fermi liquid?}
\label{sec_marginal}
Because of the smallness of the ratio $v_F/c$ and for simplicity
the Coulomb interactions in graphene are often modeled with a static scalar field that couples only the the charge density. This model was analyzed in
\cite{GGV99}.  The original motivation of this work was to
explore if the gapless free infrared fixed point encountered in
\cite{GGV94} within the perturbative RG scheme would survive in
the simplest non-perturbative resummation consisting in computing the one loop electron self--energy with an effective interaction given by the bubble sum in the photon propagator (1/N approximation).  The logarithmic renormalization of the electron wave function found in this case has led to some confusion on the marginal Fermi liquid nature of the system that it is worth to clarify.

The so--called marginal Fermi liquid (MFL)  behavior \cite{Vetal89} was set as a phenomenological model to explain some anomalous  behavior of the cuprates  high-$T_c$ superconductors. It has two
prominent features: a linear behavior of the scattering rate with
the energy near the Fermi surface
\begin{equation}
\Sigma({\bf k}, \omega)\sim g^2N^2(0)[\omega\frac{\log
x}{\omega_c}-i\frac{\pi}{2}x], \label{marginalself}
\end{equation}
where x=max$[\vert\omega\vert, T]$ , and, more important, a
vanishing quasiparticle residue at the Fermi energy:
\begin{equation}
G({\bf k}, \omega)=\frac{1}{\omega-\varepsilon_k-\Sigma({\bf
k},\omega)}=\frac{Z_k}{\omega-E_k+i\Gamma_k}+G_{{\rm incoh}},
\end{equation}
\begin{equation}
Z_k^{-1}=1-\frac{\partial\Re\Sigma(\omega)}{\partial\omega}\vert_{\omega=E_k}
\sim\log\vert\frac{\omega_c}{E_k}|.
\end{equation}
The two are related by causality since they are extracted from  the real and imaginary part of the electron Green's function.
The quasiparticle weight  $Z_k$ vanishes logarithmically at the
Fermi surface $E_k=0$ and the Green's function is entirely
incoherent. As pointed out in the original MFL reference, such a
behavior is only possible if perturbation theory, starting from a
non-interacting Fermi gas, breaks down at some energy. We must
also notice that the postulated MFL behavior in the high-$T_C$
superconductors is expected to arise from the strongly correlated
nature of the  electron--electron interaction in these compounds.
In the case of graphene the linear scattering rate comes from the
singular nature of the unscreened Coulomb interaction ($N(0)=0$ in
eq. (\ref{marginalself})).

The (instantaneous) Coulomb interaction in graphene is described by the
scalar component of the gauge field:
\begin{equation}
{\cal H}_{\rm int}= e\int d^2 {\bf r} \bar{\Psi}({\bf r})
 \Psi ({\bf r})\phi({\bf r})\;, \label{hscalar}
\end{equation}
where the scalar field has the propagator
\begin{equation}
\phi({\bf r_1}, t_1; {\bf r_2}, t_2)=\frac{1}{4\pi}
\delta(t_1-t_2)\frac{1}{\vert{\bf r_1}-{\bf r_2}\vert}\;.
\label{scalarprop}
\end{equation}
Inserting the RPA sum in the electron propagator we get the
divergent contributions to the electron self-energy from which we
can extract the following RG equations\footnote{Notice that there
is an erratum in equation (10) of ref. \cite{GGV99} where the
derivative of $\log Z$ appears instead of Z.}
\begin{equation}
\frac{\partial Z_\Psi}{\partial\log \Lambda}=
-\frac{8}{\pi^2}(2+\frac{2-g^2}{g}\frac{{\rm arc} \cos
g}{\sqrt{1-g^2}}) +\frac{8}{\pi}\frac{1}{g},
\end{equation}
\begin{equation}
\frac{\partial v_F}{\partial\log \Lambda }=
-\frac{8}{\pi^2}v_F(1+\frac{{\rm arc} \cos g}{g\sqrt{1-g^2}})
+\frac{4}{\pi}\frac{v_F}{g},
\end{equation}
where $\Lambda$ is the ultraviolet cutoff and $g$ is proportional
to the effective coupling constant $g=\frac{e^2}{16 v_F}$. The
simultaneous resolution of these equations shows that the quasiparticle residue
goes asymptotically to a non-zero constant value \cite{GGV99}.

The infrared behavior of the  system is that of a strange Fermi liquid: although
the scattering rate grows linearly with the energy, the wave
function renormalization Z runs to  a finite value at the Fermi
point. This paradoxical behavior  can be seen as another manifestation of the ``incompleteness" of the static model which can not be used to establish the infrared nature of the system.

\section{Other RG aspects}
\subsection{Disorder and interactions}
\label{sec_disorder}
In graphene many classes of lattice defects can be
described by gauge fields coupled to the two dimensional Dirac
equation \cite{VKG10}. The standard techniques of disordered
electrons \cite{AS06} can be applied to rippled graphene by
averaging over the random effective gauge fields induced by
curvature or elastic deformations. A random distribution of defects leads to a random gauge
field, with variance related to the type of defect and its
concentration. There is an extensive literature on the problem, as
the model is also relevant to Fractional Quantum Hall states \cite{LFetal94} and
to disorder in d-wave superconductors. A random field, when
treated perturbatively, leads to a renormalization of the  Fermi velocity which 
makes it to decrease at lower energies opposing the upward renormalization induced
by the long range Coulomb interaction. The simultaneous presence of interaction and disorder
gives rise to new interesting fixed points. The issue was analyzed in \cite{SGV05}. The most interesting case arises when considering a random gauge potential which models elastic distortions and some topological defects.  There is  a line of fixed points with 
Luttinger-like behavior for each disorder correlation strength
$\Delta$ given by $v_F^*=2e^2/\Delta$. An extensive analysis of the issue in the large N limit is done in \cite{FA08}.

\subsection{Short range interactions.}

As we have seen short range interactions, such an onsite
Hubbard term $U$ are irrelevant, a fact due to the vanishing density of states at the Fermi level. 
As mentioned above, the density of states at low
energies is increased by the presence of disorder. This, in turn,
enhances the effect of short range interactions that were discussed in the early work  
\cite{GGV01} and have attracted some attention recently. These interactions can be relevant in the strong coupling regime of the hexagonal electronic and optical lattices \cite{ST92,LMC01,Petal05}.

A summary of the situation following  \cite{GGV01} is as folllows: 
In the absence of disorder, an onsite Hubbard term
favors antiferromagnetism. An antiferromagnetic phase, however, is
likely to be suppressed by disorder, especially by the presence of
odd numbered rings in the lattice. Then, the next leading
instability that such an interaction can induce is towards a
ferromagnetic phase.

If a magnetic phase does not appear, electron electron
interactions, even when they are repulsive, will lead to an
anisotropic ground state. The existence of two inequivalent Fermi
points in the Brillouin zone suggests that the superconducting
order parameter induced by a repulsive interaction will have
opposite sign at each point. The corresponding symmetry is p-wave.
Disorder, in addition to the enhancement of the density of
states mentioned already, will lead to pair breaking effects in an
anisotropic superconducting phase.

\subsection{Bilayer graphene}
\label{sec_bilayer}

\begin{figure}
\begin{center}
\includegraphics[height=4cm]{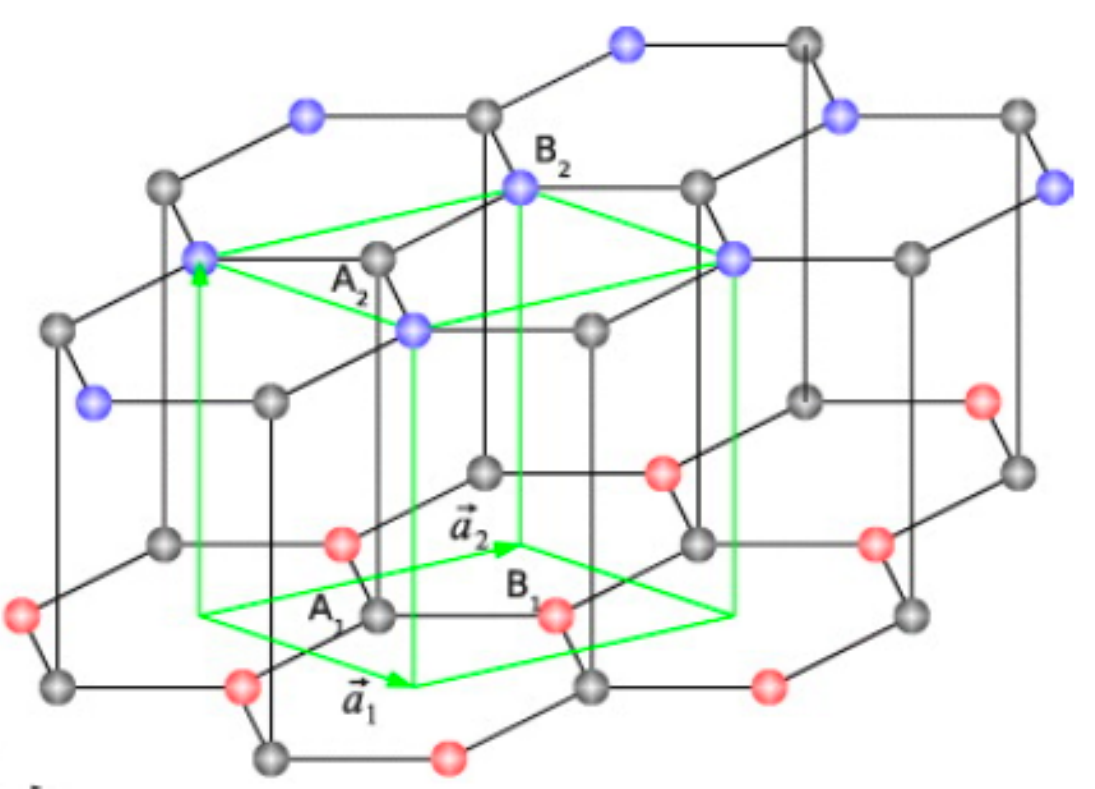}
\caption{Bilayer graphene lattice.} 
\label{fig_bilayer}
\end{center}
\end{figure}

The synthesis characterization and  analysis of the bilayer material (BG) occurred simultaneously with its monolayer counterpart \cite{Netal06}. The band structure was also theoretically described in the early publication \cite{SW58}. The most common structure is made of two graphene layers arranged in the so--called Bernal stacking as shown in Fig. \ref{fig_bilayer}. It was soon realized that the BG was even more promising than the monolayer due to the possibility of opening and controlling a gap in the system\cite{CNetal07,Oetal08}. Under a QFT point of view it constitutes an interesting example of a chiral system with quadratic dispersion relation.

A tight binding approach done to the minimal Bernal stacking bilayer shows that the system has four bands: Two low energy bands touching at a Fermi point and two at higher energy. An effective low energy continuum model \cite{MF06} keeping only the lower bands can be described by the Hamiltonian

\begin{equation}
\label{bieff} \mathcal{H}^{eff}=\sim \left(
\begin{array}{cc}
0 & k^{*2}\\
k^{2} & 0
 \end{array}
\right) ,
\end{equation}
where $k=(k_x+ik_y)$.  The dispersion relation is quadratic but
still has a non-trivial chirality.  Unlike the free monolayer
system whose effective description corresponds to regular massless
fermions, there is no QFT with quadratic propagator and Dirac
structure. In this sense, although being chiral, the Fermi points of the bilayer can not
be termed as Dirac points.
The interacting model has been subjected
to a great interest recently due to the 
enhanced quality of the samples \cite{FMetal09,WAetal10}. 

The role of  interactions and possible instabilities in the bilayer system was first analyzed in \cite{NCetal06} and has received renewed attention more recently  \cite{Xetal10,XC10,AC09,GGM10}.  RG studies have  been performed in
\cite{BY09,VY10,NL10}. The interest of this case relies on the fact that while the quadratic
dispersion makes short range interactions marginal, the existence of Fermi points (although 
with finite density of states) can give rise to departures from the Fermi
liquid behavior. Marginal Fermi liquid behavior
similar to the one discussed for the monolayer in 
Sect. \ref{sec_marginal}  has been advocated in \cite{BY09} and a fairly 
complete list of magnetic and pairing instabilities has been described in \cite{NL10,V10}. 

\section{An attempt of a summary}
\label{sec_conclusions}
The infrared properties of graphene described in this work explain the success of the non-interacting model to describe most of the low energy electronic and transport properties of the system. We have seen that the only interactions that can play a role in the system are of the form of gauge couplings. Of these, the most important is the unscreened Coulomb interaction that runs in the infrared to a very weakly interacting fixed point. In the  non-relativistic model the interaction is  marginally irrelevant. 

As a very interesting theoretical issue that makes the model differ from both  its QFT  and its condensed matter counterparts is the fact that the coupling constant renormalization at the one loop level is due to the upward renormalization to the Fermi velocity to the infrared. This very fact sets a lower bound on the validity of the widely used instantaneous model for the Coulomb interactions which ceases to be valid when the Fermi velocity approaches the speed of light. Although this bound is so small that can be ignored for most practical matters \cite{JGV10} it is conceptually important that the infrared nature of the system has to be decided with the full retarded model. 

Another interesting issue is that the running of the Fermi velocity stops at a non-trivial fixed point where its value equals the speed of light so that the Lorentz covariance appears as an emergent property \cite{V03}. Unlike most infrared stable QFT where the fixed point is trivial (g=0), the graphene system has a non-trivial fixed point characterized by the fine structure constant.

Within the static approximation a two loops calculation of the electron inverse lifetime shows that it is linear in the energy, a behavior that persists in the 1/N approximation and that is anomalous for a metallic system.

Experimental indications of both the behavior of the inverse
lifetime of the electron linear in the energy \cite{LLA09,Jetal07} and of the Fermi
velocity renormalization \cite{Letal08} have been
recently reported. The irrelevance of the short range interactions
excludes a priori any intrinsic superconductivity or magnetism in
the samples. The RG results also exclude the possibility of
opening a gap in the weak coupling regime, a fatal situation for
the electronic applications. The possibility of opening a gap in
the spectrum with or without the help of a magnetic field is
crucial for the possible electronic applications and was the main
concern in the strong coupling analysis. A recent update of the
issue and the main references can be found in \cite{K09}. A non--perturbative RG analysis of the system in the lattice has been done in \cite{GMP10}. 

\section{Acknowledgments}
I thank A. Cortijo, A. G. Grushin, F. de Juan, and B. Valenzuela for very useful
conversations and critical reading of the manuscript.
Support by MEC (Spain) through grant FIS2008-00124 is acknowledged.

\bibliography{Graphene}
\end{document}